# Light-induced nuclear quadrupolar relaxation in semiconductors

D. Paget,[a*] T. Amand,[b] J.-P. Korb[a]

[a] Laboratoire de Physique de la Matière Condensée, Ecole Polytechnique, CNRS, 91128 Palaiseau Cedex, France.

[b] Laboratoire de Physique et Chimie des Nano-Objets, INSA-CNRS-UPS, 135 avenue de Rangueil, 31077 Toulouse Cedex 4, France.

Abstract:

Light excitation of a semiconductor, known to dynamically-polarize the nuclear spins by hyperfine contact interaction with the photoelectrons, also generates an intrinsic nuclear *depolarization* mechanism. This novel relaxation process arises from the modulation of the nuclear quadrupolar Hamiltonian by photoelectron trapping and recombination at nearby localized states. For nuclei near shallow donors, the usual diffusion radius is replaced by a smaller, quadrupolar, radius. If the light excitation conditions correspond to partial donor occupancy by photoelectrons, the nuclear field experienced by electrons trapped at shallow donors can be decreased by more than one order of magnitude.

* Corresponding author: daniel.paget@polytechnique.fr

I Introduction

In a semiconductor, the possibility to enhance the nuclear polarization by the hyperfine contact interaction with spin-polarized electrons generated by circularly-polarized light excitation is of interest both for fundamental reasons and, among others, for applications to: i) quantum computing,[1] ii) transfer of nuclear magnetization to biological systems, as an alternative to adsorption of polarized xenon,[2,3] iii) understanding of the fractional quantum Hall effect.[4] Further potential applications of the optical increase of NMR sensitivity include extension to nuclei of *single spin* investigations using magnetic resonance force microscopy at surfaces.[5]

After the demonstration of optical nuclear polarization in silicon,[6] a number of recent investigations of the optically-enhanced *bulk* nuclear magnetization have been undertaken using standard NMR in Si,[7] GaAs,[8-13] InP,[14] CdTe.[15] Some of the results[11-13] were used to verify the predictions of a general theory for nuclear relaxation in solids according to which the presence of paramagnetic impurities, or localized centers, is crucial for relaxation of the nuclear spin system.[16-18] Nuclei close to the centers are relaxed by the hyperfine interaction with the spin-polarized photoelectrons trapped at these impurities, while the bulk nuclear spin system is relaxed by spin diffusion from the latter minority nuclei. A diffusion radius is defined corresponding to the distance from the impurity separating the two types of relaxation processes.[19]

Optical detection of NMR, from the depolarization at resonance of the luminescence, was first reported for GaAlAs in 1974,[20] and subsequently applied to several III-V semiconductors,[21-24] as well as 2D systems[8,25] and quantum dots.[26,27] For bulk materials, this technique was shown to only detect nuclei near the sites of electronic localization, which verifies the existence of a diffusion radius.[28] The ratio of the nuclear hyperfine field acting on the electrons and of the optically measured electronic spin polarization is consistently smaller than its calculated value. The corresponding reduction of the nuclear field is found to be of 0.1 for GaAs,[29] 0.02 for GaSb,[22] and of several percent for InP.[23] Such decrease is likely to significantly reduce the optical enhancement of the nuclear polarization.

The identification of the relaxation mechanisms responsible for this loss of nuclear polarization remains an open problem. In the absence of light excitation, the hyperfine coupling with the unpolarized holes,[30] or the quadrupolar interaction modulated by lattice phonons[31] are negligible at low temperature. The total hyperfine field of nuclei near shallow donors is decreased because of the competition between spin-lattice relaxation and spin diffusion, but only by a factor 3.[28] Another possibility is that the averaging caused by spin exchange between trapped electrons and free electrons reduces the effective nuclear field measured experimentally.[21] Interestingly, in addition to the dynamic nuclear polarization, light excitation also creates an *intrinsic leakage mechanism* for the same nuclei as the ones which are dynamically-polarized. The nuclei close to shallow donors experience a very strong electric field from the ionized donor. Since the latter field is modulated by trapping and recombination of photoelectrons, there results a significant nuclear depolarization.

The present work is devoted to an evaluation of the efficiency of such light-induced nuclear relaxation for the case of nuclei near shallow donors. In Sec. II, the characteristic time of the quadrupolar-induced evolution of the nuclear spin temperature is calculated using the semi-classical rate equation for the nuclear spin density matrix.[32] Quantitative estimates of the nuclear magnetization as a function of distance to the donor and of the nuclear field experienced by electrons trapped at shallow donors are performed in Sec. III, using the known magnitudes of quadrupolar[33-36] and hyperfine couplings.[29] Provided the light power density is such that shallow donors are partially occupied, the light-induced quadrupolar relaxation is found to induce a decrease of the nuclear hyperfine field by as much as one order of magnitude. The corresponding effect in quantum dots and the resulting dependence of the nuclear field as a function of temperature and light excitation power will be discussed elsewhere.[37]

II Light-induced quadrupolar nuclear relaxation time and nuclear polarization value

In the absence of a trapped photoelectron, the electric field experienced by nuclei near a shallow donor is given by

$$E_{off}(r) = \frac{|e|}{4\pi\varepsilon\varepsilon_0} \cdot \frac{1}{r^2} \quad (1)$$

where $e$ is the electronic charge, $\varepsilon$ is the static dielectric constant and $r$ is the distance from the donor. Photoelectron trapping and recombination induces a modulation of the electric field between Eq. (1) and $E_{on}(r)$ such that

$$E_{on}(r) = E_{off}(r)\left[1 - s(r)\right] \quad (2)$$

where the expression for $s(r)$, found using Gauss's theorem and the shape of the electronic wavefunction, is

$$s(r) = 1 - \left[1 + \frac{2r}{a_0^*} + \frac{2r^2}{a_0^{*2}}\right]e^{-2r/a_0^*} \quad (3)$$

Here $a_0^*$ is the electronic Bohr radius. In GaAs, one has $s(a_0^*) \approx 0.3$ and $E_{off}(a_0^*)$ is of the order of $10^6$V/m. The modulation amplitude $E_{off}(r) - E_{on}(r)$ induced by photoelectron trapping and recombination is very large. Unlike the usual quadrupolar relaxation, the corresponding relaxation process does not rely on phonons for modulation and can be relevant at low temperature. The present section is devoted to the calculation of the corresponding relaxation time and of the resulting decrease of the nuclear polarization.

A Quadrupolar Hamiltonian

The nuclear spin Hamiltonian, given by $H = Z + H_{IS} + H_{SS} + H_Q$, is the sum of the Zeeman term $Z$, of the hyperfine Hamiltonian $H_{IS}$, of the nuclear spin-spin interaction $H_{SS}$ and of the quadrupolar interaction $H_Q$. The expressions for the first three terms can be found in Ref. (29). For a cubic semiconductor, the expression for the quadrupolar Hamiltonian is given in Appendix A for arbitrary magnetic field B and sample surface orientations. If the magnetic field is perpendicular to a (001) sample surface, which is the case of a wide majority of experimental situations, the quadrupolar Hamiltonian is simpler:

$$H_Q = F_Q(r) . \sum_{k=1}^{2} \left[ A_{Qk} + A_{Qk}^+ \right] \tag{4}$$

Taking the normal z to the surface as the quantization axis, the spin operators $A_{Qk}$ are given by

$$\begin{aligned} A_{Q1} &= \sin\theta \, e^{i(\varphi-\pi/2)} \left[ I_z I_+ + I_+ I_z \right] \\ A_{Q2} &= -i\cos\theta \, I_+^2 \end{aligned} \tag{5}$$

and the Hermitian conjugate operators $A_{Qk}^+$ are obtained by replacing i by –i and therefore $I_+$ by $I_-$. Here $\theta$ is the angle between z and the direction Z of the electric field $\vec{E}$, and $\varphi$ is the angle between the x direction and the zZ plane. The operators $A_{Qk}$ induce transitions between Zeeman spin levels separated by energies given by

$$\hbar\omega_k = k\hbar\gamma B \quad (k=1,2) \tag{6}$$

where γ is the nuclear gyromagnetic ratio. It is convenient to write [38,39]

$$F_Q(r) = \frac{eR_{14}Q}{4I(2I-1)} E(r) = \hbar\gamma b_Q \, E(r) \tag{7}$$

where $e$ is the electronic charge, $Q$ is the quadrupolar moment of the bare nucleus of spin $I$. The factor $R_{14}$, which includes the electrostatic antishielding, is in the present frame of coordinates *Oxyz* the value of the only nonzero components of the third rank tensor relating the electric field gradient to the electric field.[33-36] The quantity $b_Q = eR_{14}Q[4\hbar\gamma I(2I-1)]^{-1}$ is the ratio of a magnetic to an electric field. It is calculated in Appendix A for different compounds and is given in Table I. The Hamiltonian $H_Q$ can be rewritten as the sum of a static and of a modulated part

$$H_Q = [1+h(t)]F_{0Q}(r).\sum(A_{Qk} + A_{Qk}^+) \qquad (8)$$

where $F_{0Q}(r)$ is given by

$$F_{0Q}(r) = [1-s(r)\Gamma_t].F_{Q\,off}(r) = \hbar\gamma[1-s(r)\Gamma_t]b_Q\,E_{off}(r) \qquad (9)$$

and $\Gamma_t$ is the fraction of the time during which the electron is present at the donor site. The function $h(t)$ describes temporal fluctuations due to the trapping and recombination of an electron at the localized site. This function has a time average equal to zero and varies randomly between $s(r)\Gamma_t[1-s(r)\Gamma_t]^{-1}$ and $-s(r)(1-\Gamma_t)[1-s(r)\Gamma_t]^{-1}$. Its correlation function, as found in Appendix B, is given by

$$g(\tau) = <h(t)h(t-\tau)> = \frac{\Gamma_t(1-\Gamma_t)s(r)^2}{[1-s(r)\Gamma_t]^2}\cdot e^{-|\tau|/\tau_{cQ}} \qquad (10)$$

The latter result expresses the fact that the interaction is not modulated for $s = 0$ or $\Gamma_t = 0$ or $\Gamma_t = 1$. The correlation time $\tau_{cQ}$ for the quadrupolar interaction is the sum of two independent contributions

$$1/\tau_{cQ} = 1/\tau_r + 1/\tau_c \qquad (11)$$

where $\tau_r$ is the recombination time of the electron trapped at the donor and $\tau_c$ is the lifetime of the ionized donor due to capture of a free electron.

B Calculation of the nuclear relaxation time

Following a semi-classical treatment, the quadrupolar-induced evolution of the nuclear spin density matrix $\sigma^*$ for the nuclear spin system, in the interaction representation and within the secular approximation, is given by[32]

$$\left.\frac{d\sigma^*}{dt}\right|_Q = -\frac{i}{\hbar}[H_0, \sigma^*] - \frac{[F_{0Q}(r)]^2}{\hbar^2} \sum_k [A_{Qk}, [A_{Qk}^+, \sigma^* - \sigma_0]] J_Q(\omega_k) \quad (12)$$

where $H_0$ is the total static Hamiltonian and $\sigma_0$ is the steady-state value of $\sigma^*$. The spectral density function $J_Q(\omega_k)$, taken for $\omega_k$ defined by Eq. (6), is given by[40]

$$J_Q(\omega_k) = \int_{-\infty}^{+\infty} e^{-i\omega_k \tau} g(\tau) d\tau \quad (13)$$

Here, we assume that the existence of interactions between nuclei results in the establishment of a temperature among the nuclear spin system. With the latter hypothesis, justified in Sec. IIID, the nuclear spin density matrix is, in the high temperature limit, of the form[29]

$$\sigma \approx [1 - \beta(Z + H_{IS} + H_Q + H_{SS})]/Tr(1) \quad (14)$$

where $\beta = 1/k_B T_n$, $k_B$ is the Boltzmann constant and $T_n$ is the temperature of the nuclear spin system. It is then found that the nuclear mean spin lies along the direction of the magnetic field independently on the relative magnitudes of Zeeman and quadrupolar interactions.

Since the latter operator $\sigma$ commutes with the static Hamiltonian, the density matrix in the interaction representation is $\sigma^* = \sigma$ and also the first term of Eq. (12) vanishes. An equation for evolution of the inverse nuclear spin temperature β is obtained, after multiplication of Eq. (12) by $I_z$, taking the trace, and using Eq. (14). Assuming that $\sigma \approx (1 - \beta Z)/Tr(1)$ [these large magnetic field conditions are defined more precisely in Sec. IIID], one obtains

$$\left.\frac{\partial \beta}{\partial t}\right|_Q = -\frac{1}{\hbar^2}\left[F_{0Q}(r)s(r)\right]^2 \frac{\Gamma_t(1-\Gamma_t)}{\left[1-s(r)\Gamma_t\right]^2}\left[\sum_k \frac{2K_k(\theta)\tau_{cQ}}{1+\omega_k^2\tau_{cQ}^2}\right](\beta-\beta_L) \quad (15)$$

Here $\beta_L = 1/k_B T_L$, $T_L$ being the temperature of the lattice. The numerical, angle-dependent, quantity $K_k(\theta)$, defined by

$$K_k(\theta) = Tr\left\{I_z\left[A_{Qk},\left[A_{Qk}^+,I_z\right]\right]\right\}/Tr\left[I_z^2\right] \quad (16)$$

is calculated in Appendix C. Its value is as expected zero for $I = \frac{1}{2}$ and is given by

$$K_1(\theta) = \frac{2}{5}[4I(I+1)-3]\frac{E_{off\perp}^2(r,\theta)}{E_{off}^2(r)} \quad (17)$$

$$K_1(\theta) + K_2(\theta) = \frac{2}{5}[4I(I+1)-3]\cdot\left[1+3\frac{E_{off//}^2(r,\theta)}{E_{off}^2(r)}\right] \quad (18)$$

where we recall that the parallel and perpendicular components of the electric field, defined with respect to the normal z to the surface, are equal to $E_{off}\cos\theta$ and $E_{off}\sin\theta$, respectively. The quadrupolar relaxation rate is finally given by

$$\frac{1}{T_{1Q}(r,\theta)} = \Gamma_t(1-\Gamma_t)\left[\gamma\, b_Q\left[E_{off}(r)-E_{on}(r)\right]\right]^2\cdot\left[\frac{2K_1(\theta)\tau_{cQ}}{1+\omega_1^2\tau_{cQ}^2}+\frac{2K_2(\theta)\tau_{cQ}}{1+\omega_2^2\tau_{cQ}^2}\right] \quad (19)$$

Its value is proportional to the square of the amplitude of the modulated electric field and further depends on the angle $\theta$ which defines the direction of the electric field.

C Steady-state nuclear mean spin

The rate equation for the evolution of the nuclear mean spin along z, neglecting the thermodynamic nuclear and electronic polarizations in the applied magnetic field as well as nuclear spin lattice relaxation processes other than the hyperfine and quadrupolar ones, is given by

$$\frac{d<I_z(r,\theta)>}{dt} = -\frac{1}{T_{1H}(r)}\left(<I_z(r,\theta)> -\frac{4}{3}[I(I+1)]\langle S_z\rangle\right)-\frac{1}{T_{1Q}(r,\theta)}<I_z(r,\theta)>+D\Delta<I_z(r,\theta)> \quad (20)$$

where $T_{1H}(\vec{r})$ is the relaxation time due to the hyperfine coupling and $<S_z>$ is the mean electronic spin along the direction z of light excitation. The third term of the latter equation describes spin diffusion, due to flip-flops between neighboring spins.[10, 16-18, 28] Here $\Delta$ is the Laplacian operator and D is the diffusion constant. Throughout the present work, it will be considered that the duration of light excitation, although sufficient to polarize the nuclei close to the donor by spin-lattice relaxation, is too short to allow this polarization to be transferred to the bulk nuclei by spin diffusion. As shown in Sec. IIIC, in the latter case, spin diffusion only marginally modifies the results of the present section so that this term will not be considered here. The steady-state value of the nuclear mean spin under the sole effect of spin-lattice relaxation is given by

$$\langle I_z(r,\theta) \rangle = p(r,\theta) \frac{4}{3}[I(I+1)]\langle S_z \rangle = \frac{4}{3} \frac{f(r,\theta)}{1+f(r,\theta)}[I(I+1)]\langle S_z \rangle \qquad (21)$$

where $0 < p(r,\theta) < 1$ expresses the reduction of nuclear mean spin with respect to its maximum value $\frac{4}{3}[I(I+1)]\langle S_z \rangle$. The quantity $f(r,\theta)$, given by

$$f(r,\theta) = \frac{T_{1Q}(r,\theta)}{T_{1H}(r)} \qquad (22)$$

is equal to $p(r,\theta)$ in the extreme case where the quadrupolar relaxation is much more efficient than the hyperfine one. The relaxation time $T_{1H}(r)$ is given by[28]

$$\frac{1}{T_{1H}(r)} = \Gamma_t \left[\gamma b_e^*(r)\right]^2 \frac{2\tau_{cH}}{1+\omega_H^2 \tau_{cH}^2} \qquad (23)$$

Here, $b_e^*(r)$ is the *instant* electronic hyperfine field acting on the nuclei. The time $\tau_{cH}$ is the correlation time of the hyperfine interaction. The energy $\hbar\omega_H$, corresponding to the flip-flop of an electronic and a nuclear spin, is given by

$$\hbar\omega_H \approx \hbar\gamma_e (B \pm B_n) \qquad (24)$$

where $B_n$ is the nuclear hyperfine field acting on the electrons, which is added or subtracted to B depending on the sign of the electronic spin. The latter energy, which depends on the *electronic* gyromagnetic ratio $\gamma_e$, is larger than $\hbar\omega_1$ and $\hbar\omega_2$ by about three orders of magnitude.

Assuming that $\omega_H^2 \tau_{cH}^2$, $\omega_1^2 \tau_{cQ}^2$, and $\omega_2^2 \tau_{cQ}^2$ are small with respect to unity, which sets an upper limit to the magnetic field value, the quantity $f(r, \theta)$ is finally given by

$$f(r,\theta) \approx \frac{\tau_{cH}}{\tau_{cQ}} \cdot \frac{1}{(1-\Gamma_t)} \left[ \frac{b_e^*(r)}{b_Q [E_{off}(r) - E_{on}(r)]} \right]^2 \cdot \left[ \sum_k K_k(\theta) \right]^{-1} \quad (25)$$

Note that, since the spatial dependence of the electric fields $E_{off}(r)$ and $E_{on}(r)$ does not appear explicitly, Eq. (19) and Eq. (25) are valid for any localized electronic state. For nuclei near a donor one has

$$b_Q E_{off}(r) = b_Q E_{off}(a_0^*) \cdot (a_0^*/r)^2 \quad (26)$$

$$b_e^*(r) = b_e^*(a_0^*) e^{-2(r/a_0^* - 1)} \quad (27)$$

Using Eq. (2), Eq. (17), and Eq. (18), it is possible to separate $f(r,\theta)$ into the product of a radial dependence $\varphi(r)$, of an angular one, and of a numerical coefficient $f_0$ which is a measure of the relative strengths of hyperfine to quadrupolar relaxations:

$$f(r,\theta) \approx \frac{f_0 \varphi(r)}{1 + 3\cos^2\theta} \quad (28)$$

$$\varphi(r) = \frac{e^{-4(r/a_0^* - 1)}}{s(r)^2} \left( \frac{r}{a_0^*} \right)^4 \quad (29)$$

$$f_0 = \frac{5}{2} [4I(I+1) - 3]^{-1} (1 - \Gamma_t)^{-1} \frac{\tau_{cH}}{\tau_{cQ}} \left( \frac{b_e^*(a_0^*)}{b_Q E_{off}(a_0^*)} \right)^2 \quad (30)$$

The implications of the latter equations are discussed in the following section.

III Discussion

A Effect of the donor rate of occupation.

A key parameter for the value of $f_0$ is the rate of occupation of the donors $\Gamma_t$ by photoelectrons, which depends on the light excitation power. Indeed:

a) The correlation times $\tau_{cH}$, and $\tau_{cQ}$ depend on the free electron density $n_f$. The time $\tau_{cQ}$ can be written using Eq. (11)

$$\frac{1}{\tau_{cQ}} = \frac{1}{\tau_r} + \sigma_c v n_f \tag{31}$$

where $v$ is the velocity of free electrons and $\sigma_c$ is the cross section for their capture at donors. The correlation time $\tau_{cH}$ of the hyperfine interaction is given by

$$\frac{1}{\tau_{cH}} = \frac{1}{2\tau_r} + \frac{1}{T_1} + \frac{1}{\tau_{ex}} \approx \sigma_e v n_f \tag{32}$$

as obtained in Appendix B, assuming that the electronic polarization is weak with respect to unity. Here $T_1$ is the electronic spin–lattice relaxation time and $\tau_{ex}$ is the characteristic time for spin exchange between trapped and free electrons. In GaAs, for above bandgap light excitation, it has been found that the latter process is dominant by several orders of magnitude, so that $\tau_{cH}$ has a simple approximate expression, also given in Eq. (32), where $\sigma_e$ is the spin exchange cross section.[21]

b) The rate $\Gamma_t$ of donor occupation is obtained by writing the rate equation for the population of electrons trapped at donors, of concentration $N_D$. The latter equation, given in Appendix D considering above bandgap light excitation, yields

$$\Gamma_t = \frac{\sigma_c \tau_r v n_f}{1 + \sigma_c \tau_r v n_f}. \tag{33}$$

Using Eq. (28), Eq. (29) and Eq. (30), $f_0$ is given by

$$f_0 = \frac{5}{2}\frac{\sigma_c}{\sigma_e}[4I(I+1)-3]^{-1}\left(\frac{b_e^*(a_0^*)}{b_Q E_{off}(a_0^*)}\right)^2 \frac{1}{\Gamma_t(1-\Gamma_t)} = \frac{f_{00}}{\Gamma_t(1-\Gamma_t)} \quad (34)$$

Eq. (34) has a simple form in which the quantity $f_{00}$, which is a measure of the maximum magnitude of the quadrupolar–induced loss of nuclear magnetization, is independent of experimental conditions such as excitation power. The latter dependence is concentrated in the donor occupation rate $\Gamma_t$. According to Eq. (34), the quadrupolar-induced loss of nuclear polarization occurs when the donors are *partially* occupied, which can be easily characterized from the power dependence of the donor luminescence. For a density of conduction electrons much smaller than $(\sigma_c \tau_r v)^{-1}$, one has $\Gamma_t << 1$ and the quadrupolar effects are small since the correlation time $\tau_{cH}$ is large. Conversely, if $\Gamma_t = 1$, the quadrupolar interaction is not modulated and cannot relax the nuclear spins.

B Order of magnitude estimates

For $As^{75}$ in GaAs, the efficiency of the quadrupolar relaxation process comes from the fact that the spin exchange cross section [$\sigma_e \sim 9 \times 10^{-16}$ m$^2$] [21] is three orders of magnitude larger than the one for electron capture at donors [$\sigma_c = 5.1 \times 10^{-19}$ m$^2$].[41] Using Table I and Ref. (29), we obtain $b_e^*(a_0^*) = 1.5 mT \approx 5.2 b_Q E_{off}(a_0^*)$ and we find $f_{00} \sim 2 \times 10^{-3}$, and $f_0 \sim 10^{-2}$ for $\Gamma_t = 1/2$. As found from Eq. (28), $f(a_0^*, \pi/2) \approx 10^{-1}$ and $f(a_0^*, 0) \approx 2 \times 10^{-2}$, so that the nuclei at the Bohr radius are depolarized by the quadrupolar relaxation.

Nuclei such as $In^{115}$ in InP and $Sb^{121}$ in GaSb are believed to exhibit stronger quadrupolar effects because of their larger spin values (9/2 for $In^{115}$ and 5/2 for $Sb^{121}$). However, as seen in Table I, the quantity $b_Q$ is smaller than for $As^{75}$ in GaAs. Using Table I and Eq. (34) and assuming that both $\sigma_c$ and $\sigma_e$ scale like the Bohr radius, so that their ratio is independent on material, we estimate that $f_{00}$ is equal to $9.4 \times 10^{-3}$ and $1.2 \times 10^{-2}$ for $In^{115}$ in InP and Sb$^{121}$ in GaSb respectively. This implies that the latter materials should also exhibit nuclear polarization losses of quadrupolar origin, although slightly smaller than for GaAs.

C Radial and angular dependences of the nuclear polarization: quadrupolar diffusion radius

Shown in Fig. 1 are the radial dependences of $p(r, 0)$ and $p(r, \pi/2)$ using $f_0 \sim 10^{-2}$. Close to the donor position, one has $p(r, \theta) = 1$, as the quadrupolar relaxation is inefficient because $s(r) \approx \frac{4}{3}(r/a_0^*)^3$ so that the electric field is not modulated. As a function of distance, although the quadrupolar rate first increases and then decreases, $f(r, \theta)$ exhibits a monotonic, decreasing behavior. The nuclei are depolarized above a distance to the donor corresponding to $f = 1$. As seen in Fig. 1, this distance is smaller in the direction z of the magnetic field ($0.25 a_0^*$) than in the perpendicular directions ($0.45 a_0^*$).

For calculation of the nuclear field experienced by trapped electrons, two approximations will be made. First, we shall use for simplicity the *angular* average of the nuclear polarization, defined as $< p(r) > = \int \sin\theta\, p(r,\theta) d\theta\, d\varphi / \int \sin\theta\, d\theta\, d\varphi$. As found using Eq. (21) and Eq. (28), this quantity is given by

$$< p(r) > = \frac{f_0 \varphi(r)}{\sqrt{3[1 + f_0 \varphi(r)]}} Arctg\left[\frac{\sqrt{3}}{\sqrt{1 + f_0 \varphi(r)}}\right] \quad (35)$$

for which the radial dependence, also shown in Fig. 1, is intermediate between those of p(r, 0) and p(r, π/2). The second approximation consists in replacing <p(r)> by a step function at r = $\rho_Q$ such that

$$<p(\rho_Q)> = \frac{1}{2} \quad (36)$$

The nuclear hyperfine field, defined by $B_n = B_{n0} a_0^{*-3} \int_0^\infty 4r^2 e^{-2r/a_0^*} < p(r) > dr$,[29] is approximated by

$$B_n \approx B_{n0} s(\rho_Q) \tag{37}$$

where $B_{n0}$ is the nuclear field value for a homogeneous nuclear polarization and $s(r)$ is defined in Eq. (3). The latter approximation implies that the quadrupolar relaxation is inefficient for distances smaller than $\rho_Q$ and dominant for larger distances ($f=0$). Such approximation is usual in analyses of nuclear polarizations near shallow donors,[16-18] and results in defining a sphere around the donor inside which the nuclear polarization is not affected by the quadrupolar relaxation. The radius of this sphere, which will be called the quadrupolar radius, replaces the usual diffusion radius for the estimate of the nuclear hyperfine field. Shown in Fig. 2 are the variations of $\rho_Q$ and of $s(\rho_Q)$, as a function of $f_0$. For $f_0 = 10^{-2}$, one finds $\rho_Q \sim 0.35 a_0^*$ which leads to $B_n = 0.03 B_{n0}$.

We now discuss the effect of spin diffusion between neighboring spins, which appears as the third term of Eq. (20) and has so far been neglected. A diffusion radius $\rho_D$ is defined, corresponding to the distance from the donor at which the efficiencies of direct relaxation and of spin diffusion are equal. This radius is given by

$$\frac{1}{T_{1Q}(\rho_D)} + \frac{1}{T_{1H}(\rho_D)} = \frac{D}{\rho_D^2} \tag{38}$$

At a distance smaller than $\rho_D$, all the considerations of the present work are valid, since spin diffusion does not affect the nuclear polarization. On the other hand, for nuclei situated beyond $\rho_D$, the polarization is strongly decreased because of the efficient diffusion towards the bulk nuclei. In the absence of quadrupolar relaxation, the maximum value of the reduced nuclear field is $s(\rho_D)$. Since a value $\rho_D \sim 1.4 a_0^*$ has been found,[28] we obtain $s(\rho_D) \sim 0.5$. As seen from Eq. (38), the quadrupolar relaxation results in a decrease of $\rho_D$. For $f_0=10^{-2}$ and using the value of $D$ of Ref. (28), we calculate a modified diffusion radius $\rho_D$ value close to $a_0^*$. Since the latter value is still larger than $\rho_Q$, spin diffusion is generally negligible with respect to spin-lattice relaxation for the nuclei which contribute to the nuclear field experienced by trapped electrons.[42]

Spin diffusion should however be taken account of in two extreme cases concerning the light excitation power. In the case of a very weak efficiency of the overall spin-lattice relaxation, Eq. (38) does not have a real solution so that spin diffusion becomes predominant at all distances from the donor. Since the efficiencies of both the hyperfine and quadrupolar relaxation processes are proportional to $\Gamma_t$, such case is obtained for a weak light excitation density.[for $As^{75}$ in GaAs, we estimate that this situation corresponds to a threshold characterized by *$\Gamma_t$<0.15*] This situation is outside the scope of the present work and is not considered here. Conversely, for a high light excitation power, if $\Gamma_t \sim 1$, the quadrupolar radius $\rho_Q$ increases because the quadrupolar spin-lattice relaxation becomes negligible with respect to the hyperfine one. When $\rho_Q$ is larger than $\rho_D$, since the polarization of nuclei at a distance larger than $\rho_D$ is decreased by spin diffusion towards the bulk nuclei, the nuclear field is obtained by replacing $\rho_Q$ by $\rho_D$ in Eq. (37). Thus, the maximum nuclear field obtained for a negligible quadrupolar relaxation is given by $B_{n0}s(\rho_D)$. As a result, the relative decrease of the nuclear field produced by the light-induced quadrupolar relaxation is $s(\rho_D)/s(\rho_Q)$ which, in the conditions of Fig. 1, is of the order of 15.

D Magnetic field effects

The present Subsection is devoted to the justification of three hypothesis made in Sec. II B, for which the validity depends on the magnetic field value.

1) The zero magnetic field expression of the quadrupolar-induced decrease of nuclear magnetization has been used in Eq. (25). With the values of the cross sections $\sigma_c$ and $\sigma_e$ given in Sec. III. B and taking $n_f = 10^{21} m^{-3}$, one finds that $\omega_H \tau_{cH} = 1$ for B= 20T. The same magnetic field value gives $\omega_1 \tau_{cQ} = 1$, further taking $\tau_r \sim 1$ nsec.[43] It is concluded that Eq. (25) is valid up to very large magnetic field values.

2) It has been assumed that the heat capacity of the Zeeman reservoir is larger than those of the quadrupolar and spin-spin ones. Such assumption is obviously not valid at very low magnetic field. The lower magnetic field limit is obtained by expressing the heat capacities of the various reservoirs using the following relation[29]

$$\frac{<Z>}{B^2} = \frac{<H_{ss}> + <H_Q>}{B_L^2 + B_Q^2} = -\frac{1}{k_B T_n} \frac{I(I+1)}{3} (\gamma \hbar)^2 \qquad (39)$$

where the electronic field acting on the nuclear spins has been neglected. Here $B_L$ is the local field. For a magnetic field larger than $B_L$, only flip-flops between nuclei of the same isotopic species are energetically allowed. In the particular case of $As^{75}$ in GaAs, this leads to $B_L \sim 0.03$ mT.[29] The local field of quadrupolar origin $B_Q$, equal to zero for $I = ½$, is given by

$$B_Q^2 = \frac{3 \, Tr <H_Q^2>}{I(I+1)(2I+1)(\gamma \hbar)^2} = \frac{4}{5} (b_Q E_{off})^2 (1 - s\Gamma_t)^2 [4I(I+1) - 3] \qquad (40)$$

We conclude that the high field limit discussed in Sec. IIB is valid provided

$$B^2 >> B_L^2 + B_Q^2 \qquad (41)$$

Thus the *effective* local field is larger than the spin-spin local field. For a magnetic field along the z direction, assuming for simplicity $\Gamma_t = 0$ and taking $r \approx 0.5 a_0^*$, we calculate $B_Q \sim 1.6$ mT which is more than one order of magnitude larger than $B_L$.[29]

3) The hypothesis made in Sec. IIB of a nuclear spin temperature gives an independent low magnetic field limit for the validity of the present model. In the absence of quadrupolar couplings, there is no doubt that there exists a spin temperature since the time $T_2$ of establishment of the nuclear temperature is of the order of $1/\gamma B_L \sim 300$ μsec. Inclusion of quadrupolar interactions does not change the latter picture provided the following condition is fulfilled

$$\Delta E = \left| \left[ \delta_Q E^i_m - \delta_Q E^i_{m-1} \right] - \left[ \delta_Q E^j_{m'+1} - \delta_Q E^j_{m'} \right] \right| < \hbar \gamma B_L \qquad (42)$$

where ΔE is the energy balance of a difference between transition energies, expressed as a function of the quadrupolar-induced shift $\delta_Q E^i_m$ of the level m of spin i. Here, j is the nearest neighbor of nucleus i, of the same isotopic specie. There are two distinct reasons for which the latter condition is likely not to be fulfilled:

i) For a *homogeneous* electric field, although the quadrupolar shifts of nuclei i and j are the same, one has $\delta_Q E^i_m \neq \delta_Q E^j_{m'}$ for m ≠ m'. This may prevent some flip-flops between neighboring nuclei and therefore induce a decrease of the local field. Following Abragam,[44] it is found that such effect leads to a decrease by only 15%, so that quadrupolar interactions weakly affect the time $T_2$ of establishment of a nuclear spin temperature.

ii) Near a donor, flip flops between nearest neighbors may be prohibited because of the strong spatial dependence of the electric field so that $\delta_Q E^i_m \neq \delta_Q E^j_m$ for a given quantum number m. Since the effect of distinct quadrupolar shifts of states with *distinct* m values has been examined in i) above, we replace Eq. (42) by $\left| \delta_Q E^i_m - \delta_Q E^j_m \right| < \hbar \gamma\, m B_L$, which states that, *for a fixed m*, the difference in quadrupolar shifts of neighboring nuclei is smaller than the Zeeman energy in the local field. The latter condition allows us to estimate the characteristic radius $r_Q$ of the zone outside which a spin temperature exists, using the following value of $\delta_Q E^i_m$, obtained by second order perturbation

$$\delta_Q E^i_m = \hbar \gamma\, b_Q \overline{E}(r_i) \frac{2 m b_Q \overline{E}(r_i)}{B} \left\{ \frac{\overline{E}^2_\perp (r_i, \theta)}{\overline{E}^2(r_i)} \left[ 4I(I+1) - 8m^2 - 1 \right] - \frac{\overline{E}^2_{//}(r_i, \theta)}{\overline{E}^2(r_i)} \left[ 2I(I+1) - 2m^2 - 1 \right] \right\} \qquad (43)$$

where $\overline{E}(r_i) = [1 - s(r)\Gamma_t] E_{off}(r_i)$ is the time average of the electric field. Due to the presence of the magnetic field B at the denominator of the latter equation, the radius $r_Q$ decreases with increasing magnetic field. Derivating Eq. (43) with respect to distance, we find for $r_Q$ an expression of the type:

$$r_Q = \eta B^{-1/5} \qquad (44)$$

The radius $r_Q$ is largest when the electric field is parallel to z, when $\Gamma_t << 1$ so that $\overline{E}(r_i) \approx E_{off}(r_i)$ and for m=3/2 in the case of I=3/2. In the latter case, taking account of the fact that in GaAs the i-j direction is along the [110] crystal axis, and using the known interatomic spacing, we find $\eta \approx 3 nm T^{1/5}$. It is concluded that, at a distance r from the donor, the hypothesis of nuclear spin temperature is valid provided

$$B > B'_Q = \left(\frac{\eta}{r}\right)^5 \qquad (45)$$

Since most of the effects discussed above occur for distances larger than $r = a_0^*/2$, $B'_Q$ is evaluated to ~0.09 T. Although much larger than $B_L$ and $B_Q$ defined by Eq. (41), the latter value is smaller than magnetic fields used in most experiments.

For smaller magnetic fields, the evolution of the mean nuclear spin value $<\vec{I}>$, calculated using Eq. (12) and $<\vec{I}> = Tr(\sigma \vec{I})$, is found to be nonexponential as a function of time so that the calculation of the steady-state nuclear magnetization becomes intricate. However, qualitatively, the decrease of the steady-state nuclear magnetization is still expressed by an equation of the same type as Eq. (25) with distinct numerical factors, so that the magnetization is still strongly reduced. Furthermore, because of the very weak magnetic field dependence of $r_Q$, the conclusions of the present work are still qualitatively correct for $B < B'_Q$: For a magnetic field equal to $B'_Q/3$, one has $r_Q \approx 0.6 a_0^*$. As seen using Eq. (37), such increase is of moderate impact on the nuclear field value since the relative increase of the nuclear field when $r_Q$ increases between $0.5 a_0^*$ and $0.6 a_0^*$, of 50%, is much smaller than the decreases found in the present work, of more than one order of magnitude.[45]

E Nuclear field dependence on light excitation power and doping

Here we summarize the results of the preceding subsections and obtain the value of the nuclear field $B_n$, using Eq. (37) and further considering the various processes on which depends $B_n$. The shallow acceptor concentration $N_A$ is assumed to be larger than the donor one so that the donor levels are unpopulated in the absence of light excitation. As shown in Appendix D, the donor occupancy factor $\Gamma_t$, on which depends the quadrupolar radius $\rho_Q$, is related to the excitation power density $P$ by

$$P = \frac{P_0}{(1-\xi)^2}\left[\Gamma_t + \xi \frac{N_A}{N_D}\right] \tag{46}$$

where

$$P_0 = L h \nu k N_A N_D \tag{47}$$

$$\xi = \frac{k}{\sigma_c v} \frac{\Gamma_t}{1-\Gamma_t} \tag{48}$$

Here $h\nu$ is the photon energy (assumed to be above bandgap) and $L$ is the electron diffusion length. The quantity $k$ is the coefficient for bimolecular electron-hole recombination.

Shown in Fig. 3 is the specific case of $As^{75}$ in GaAs, using for illustration purposes $N_D=10^{22}$ m$^{-3}$, $N_A=5 \times 10^{22}$ m$^{-3}$, $L \sim 5\mu m$,[46] $k \approx 1 \times 10^{-14}$ m$^3$/s,[47] and $k/\sigma_c v \approx 10^{-1}$, which corresponds to $T \sim 40K$.[41] Shown in Curve a is the dependence of $\Gamma_t$ as a function of light excitation power, obtained using Eq. (46). The light excitation power range $P < 0.2 P_0$ corresponds to $\Gamma_t < 0.15$ and is not considered here because, as discussed in Sec. IIID, spin diffusion becomes dominant. Shown in Curve b of Fig. 3 is the power dependence of the quantity $s(\rho_Q)$, obtained using Eq. (3), Eq. (34), Eq. (35), Eq. (36). Immediately apparent is the fact that, in a relatively broad power range near $P_0$, corresponding to $\Gamma_t=0.5$, the nuclear field is decreased with respect to the maximum value of 0.5 imposed by spin diffusion, by a factor of about 20. Note that, at high power, the light-induced quadrupolar relaxation still decreases the nuclear field because, as seen from Eq. (46) and Eq. (47), $\Gamma_t$, equal to $(1+k/\sigma_c v)^{-1}$, is smaller than unity. In the present case, one finds $P_0=2.4 \times 10^6 W/m^2$, a realistic value.

The *effective* nuclear field, measured experimentally from its effect on the electronic polarization, is further modified by the efficient spin exchange processes between free and trapped electrons considered in Sec. IIIA and is equal to the average of the nuclear fields experienced by the two electronic species. Such effect does not modify the quadrupolar radius, but only the multiplicative factor $B_{n0}$ defined by Eq. (37). Assuming, as performed throughout the present work, that the bulk nuclei are weakly-polarized, the nuclear field experienced by free electrons is very small so that $B_{no}$ is proportional to $\Gamma_t N_D / (n_f + \Gamma_t N_D)$ where the concentration of free electrons is given by Eq. (D6). One has finally

$$B_n = \alpha_n b_{n0} <S_0> \tag{49}$$

where $b_{n0}$ does not depend on light power and doping. Since the common mean spin $<S_0>$ of free and trapped electrons can be measured from the luminescence polarization, the quantity $\alpha_n$, given by

$$\alpha_n = \frac{\Gamma_t N_D}{n_f + \Gamma_t N_D} s(\rho_Q) \tag{50}$$

is the reduced nuclear field for which we now consider the power dependence.

Shown in Curve c of Fig. 3 is the light excitation dependence of $n_f/N_A$, obtained using Eq. (D6) and assuming $N_D=N_A/5$. Shown in Curve d is the light excitation dependence of $\alpha_n$.

For $P < P_0$, the dominant mechanism for nuclear field reduction is the light-induced quadrupolar relaxation. The nuclear field reduction due to spin exchange becomes significant for $P > P_0$ since $n_f$ increases while $\Gamma_t$ is nearly constant and induces an overall decrease of the nuclear field with light excitation power. [48]

The effect of a change of *doping* of the quadrupolar-induced reduction of nuclear field is limited to the sole variation of the quantity $P_0$. As a result, an increase of acceptor and donor doping levels simply shifts Curve b of Fig. 3 along the X axis by a similar factor without any change of shape. As seen from Eq. (50), this is still true if one includes the effect of spin exchange, provided the ratio $N_A/N_D$ remains constant. Note finally that *resonant excitation* of donor states might enable to increase the nuclear field value: As seen from Eq. (23), Eq. (32), and Eq. (50), the subsequent decrease of the free electron concentration should induce an increase of the efficiency of the hyperfine relaxation and a decrease of the effect of spin exchange.

We now discuss the possibility of experimental demonstration of the light-induced quadrupolar relaxation. In order to separate the contribution to the nuclear field value of the light-induced quadrupolar relaxation from that of spin exchange with free electrons, it is crucial to analyze the dependence of the nuclear field as a function of light excitation and donor concentration. However, among the experimental works which have estimated the leakage factor $f$,[22, 23, 29] none of them has performed the latter analysis, so that experimental proof of the present mechanism is lacking. Such analysis is beyond the scope of the present paper and will be published elsewhere for the case of quantum dots.[37]

IV Conclusion

We now summarize the main results of the present work:

a) The effect of the light-induced quadrupolar relaxation is evaluated assuming that there exists a temperature among the nuclear spin system. The latter hypothesis implies that the external magnetic field is sufficiently large to decrease the difference between the quadrupolar shifts of neighboring nuclei so that flip flops are allowed. In the latter case, the time evolution of the nuclear spin temperature is found to be exponential, so that a relaxation time can be defined. The latter time $T_{1Q}$, within numerical factors, depends on the product of the square of the modulation amplitude and of the correlation time of the modulation. Comparison of $T_{1Q}$ with the relaxation time due to the hyperfine contact interaction gives the expression for the nuclear polarization as a function of the distance to the donor under the combined effects of quadrupolar and hyperfine relaxations.

b) Near shallow donors in semiconductors, the angular-averaged effect of the quadrupolar relaxation is to replace the diffusion radius $\rho_D$ up to which the nuclei are spin-polarized by a novel, smaller, radius called the *quadrupolar radius $\rho_Q$*.

c) The quadrupolar-induced decrease of the nuclear field occurs in conditions of light excitation corresponding to partial donor occupancy by photoelectrons. This should induce a decrease of the nuclear field by more than one order of magnitude in GaAs and by slightly smaller factors for InP and GaSb. In addition, the effect of averaging of the nuclear field between free and trapped electrons, due to spin exchange, produces a further decrease of the nuclear field for larger light excitation powers.

Appendix A : Form and magnitude of the quadrupolar Hamiltonian

We consider here the general case, described by Fig. 4, where both the electric field direction Z and the magnetic field one Z' do not coincide with a crystal axis z, taken perpendicular to the crystal surface. The quadrupolar Hamiltonian $H_Q$ of a given nucleus at position $\vec{r}$ is related to the components of the electric field gradient by [38, 39]

$$H_Q(\vec{r}) = \frac{eQ}{4I(2I-1)} \left\{ \begin{array}{l} V_{Z'Z'}(\vec{r})[3I_{z'}^2 - I(I+1)] \\ + V_{X'Z'}(\vec{r})[I_{z'}(I_{+'}+I_{-'}) + (I_{+'}+I_{-'})I_{z'}] - iV_{Y'Z'}(\vec{r})[I_{z'}(I_{+'}-I_{-'}) + (I_{+'}-I_{-'})I_{z'}] \\ + \frac{1}{2}[V_{X'X'}(\vec{r}) - V_{Y'Y'}(\vec{r})][I_{+'}^2 + I_{-'}^2] - iV_{X'Y'}(\vec{r})[I_{+'}^2 - I_{-'}^2] \end{array} \right\} \quad (A1)$$

where the quantization axis Z' is the magnetic field direction, the spin operators $I_{\pm'}$ are equal to $I_{X'} \pm iI_{Y'}$ and

$$V_{ij}(\vec{r}) = \frac{\partial^2 E(\vec{r})}{\partial X'_i \partial X'_j} \quad (A2)$$

and $X'_i$ stands for $X'$, $Y'$, or $Z'$. These directions are distinct from the xyz directions of the cubic crystal lattice, z being also the normal to the sample surface. The components of the electric field gradient tensor in the $X'Y'Z'$ frame are obtained by using elementary rules for tensor transformation and are given by [35]

$$\begin{pmatrix} V_{X'X'} \\ V_{Y'Y'} \\ V_{Z'Z'} \\ V_{Y'Z'} \\ V_{X'Z'} \\ V_{X'Y'} \end{pmatrix} = R_{14} \begin{pmatrix} -\sin 2\theta' \sin\varphi' & -\sin 2\theta' \cos\varphi' & \cos^2\theta' \sin 2\varphi' \\ 0 & 0 & -\sin 2\varphi' \\ \sin 2\theta' \sin\varphi' & \sin 2\theta' \cos\varphi' & \sin^2\theta' \sin 2\varphi' \\ \cos\theta' \cos\varphi' & -\cos\theta' \sin\varphi' & \sin\theta' \cos 2\varphi' \\ \cos 2\theta' \sin\varphi' & \cos 2\theta' \cos\varphi' & \frac{1}{2}\sin 2\theta' \sin 2\varphi' \\ -\sin\theta' \cos\varphi' & \sin\theta' \sin\varphi' & \cos\theta' \cos 2\varphi' \end{pmatrix} \begin{pmatrix} E_x \\ E_y \\ E_z \end{pmatrix} \quad (A3)$$

where, as shown in Fig. 3, $\theta$ and $\varphi$ are the angles between Z and z and between x and the zZ plane, respectively, and $\theta'$ and $\varphi'$ are the angles between z and Z' and between x and the zZ' plane respectively. Here, $R_{14}$ is the sum of an ionic contribution, (which depends on the ionicity of the solid, on $\varepsilon^2$-n, where n is the infra-red optical index, and on the antishielding factor) and of the covalent contribution (which further depends on the bandgap value).[33]

The expression of the quadrupolar Hamiltonian is then obtained from Eq. (A1) and (A3). For an arbitrary orientation of the magnetic field, this expression is intricate and depends both on $\theta'$ and $\varphi'$. If the magnetic field $B$ direction coincides with a [100] crystal axis z,($\theta'=\varphi'=0$) the only nonzero components of $V_{ij}$ in the xyz frame are

$$V_{xy} = R_{14} E(\vec{r}) \cos\theta$$
$$V_{yz} = R_{14} E(\vec{r}) \sin\theta \cos\varphi \qquad (A4)$$
$$V_{zx} = R_{14} E(\vec{r}) \sin\theta \sin\varphi$$

Eq. (4) is readily obtained.

In order to estimate $b_Q$, it is necessary to determine $R_{14}$. One of the first determinations was performed for GaAs, where the effect of application of an electric field along the [111] direction on the quadrupolar splitting of the NMR line was studied.[33] Here, we take the more recent measurements of Ref. 36, which give slightly larger values, arguing that the smaller values obtained in Ref. 33 were due to sample inhomogeneities. For GaAs, InAs and GaSb, independent estimates of $R_{14}$ were obtained using the broadening of the nuclear acoustic resonance.[34] For GaAs, they differ from the latter value by about a factor of 3-4. As a result, for a nucleus $\alpha$ of InAs or GaSb, we have chosen to determine $R_{14}^{\alpha}$ according to the following scaling involving Ref. (36) and Ref. (34)

$$R_{14}^{\alpha} = R_{14}^{\alpha}(REF.34) \cdot \frac{R_{14}^{As}(REF.36)}{R_{14}^{As}(REF.34)} \qquad (A5)$$

For In[115] in InP no estimate of $R_{14}$ has to our knowledge been published. However, $R_{14}$ of In[115] in InP should not differ from that of In[115] in InAs by more than a factor of 50% since the ionicities of InAs and InP are identical and since the effect of bandgap should be similar to the ratio of the $R_{14}$ values of $As^{75}$ between GaAs and InAs. The final results are shown in Table I.

Appendix B: Correlation functions of the quadrupolar and hyperfine interactions

The modulation of the quadrupolar interaction is described by the function $h(t)$, given by Eq. (9). This function is of zero average and takes two discrete values $h_\alpha$ (where $\alpha = 1,2$) given respectively by $h_1 = s\Gamma_t(1-s\Gamma_t)^{-1}$ or $h_2 = -s(1-\Gamma_t)(1-s\Gamma_t)^{-1}$, with respective probabilities $w_1 = \Gamma_t$ and $w_2 = 1 - \Gamma_t$. The correlation function is written under the form

$$g_Q(\tau) = <h(t)h(t-\tau)> = \sum_\alpha h_\alpha w_\alpha \sum_\beta h_\beta P_{\alpha\beta}(\tau) \tag{B1}$$

where $P_{\alpha\beta}(\tau)$ is the conditional probability that $h = h_\beta$ at time $\tau$, under the condition that $h = h_\alpha$ at time $t = 0$.

Assuming that the fluctuation process is Markovian and stationary, the quantity $P_{\alpha\beta}(\tau)$ is given by [49]

$$\frac{dP_{\alpha\beta}}{dt} = \sum_\gamma \Pi(\gamma,\beta) P_{\alpha\beta}(t) \tag{B2}$$

where $\Pi(\gamma,\beta)$ is a numerical factor, equal for $\gamma \neq \beta$ to the probability per unit time that $h(t)$ goes from the value $h_\gamma$ to the value $h_\beta$. The quantity $-\Pi(\beta,\beta)$ is the probability that $h(t)$ goes from $f_\beta$ to the other value. One has $\Pi(1,2) = \tau_1^{-1}$, $\Pi(2,1) = \tau_2^{-1}$, $\Pi(2,2) = -\tau_1^{-1}$, $\Pi(1,1) = -\tau_2^{-1}$, where $\tau_\alpha$ is the lifetime of state $\alpha$ .[with the definitions of Sec. IIA, one has $\tau_1 = \tau_r$ and $\tau_2 = \tau_c$ ] Using the latter values, resolution of Eq. (B2) yields

$$\begin{aligned} P_{11} &= (1-\Gamma_t) + \Gamma_t \exp\left[-t(\tau_1^{-1} + \tau_2^{-1})\right] \\ P_{21} &= (1-\Gamma_t) - (1-\Gamma_t) \exp\left[-t(\tau_1^{-1} + \tau_2^{-1})\right] \\ P_{12} &= \Gamma_t - \Gamma_t \exp\left[-t(\tau_1^{-1} + \tau_2^{-1})\right] \\ P_{22} &= \Gamma_t + (1-\Gamma_t) \exp\left[-t(\tau_1^{-1} + \tau_2^{-1})\right] \end{aligned} \tag{B3}$$

The result of Eq. (10) is obtained after replacing $P_{\alpha\beta}$ by their latter values in Eq. (B1).

The same procedure can be applied to calculate the correlation function for the hyperfine interaction. Here three states, labeled +1, -1, or 0 are considered, depending on the absence or presence of an electron of a spin equal to +1/2 or -1/2. In addition with the recombination time $\tau_r$, the correlation time also depends on the spin-lattice relaxation time $T_1$ and of the characteristic time $\tau_{ex}$ due to possible spin-exchange processes with delocalized electrons. The final expression for the correlation function, valid in the limit of small electronic polarizations (i. e. $\tau_{ex}^{-1} + T_1^{-1} \gg \tau_r^{-1}$), is

$$g_H(\tau) = \Gamma_t \cdot e^{-|\tau|/\tau_{cH}} \tag{B4}$$

where $\tau_{cH}$ is given by Eq. (32). Eq. (B4) expresses the fact that, unlike for the quadrupolar coupling, the hyperfine relaxation is inefficient in the only case where the probability $\Gamma_t$ of occupation of the localized state is zero.

Appendix C : Expression of $K_k(\theta)$ defined by Eq. (16)

Applying the relations $Tr(ABC) = Tr(BCA)$ and $Tr\{A[B,[C,D]]\} = Tr\{[A,B][C,D]\}$ where $A$, $B$, $C$ and $D$ are spin operators, one obtains

$$Tr\{I_z[A_{Q,k},[A_{Q,k}^+,I_z]]\} = Tr\{[I_z,A_{Q,k}][A_{Q,k}^+,I_z]\} \tag{C1}$$

One finds

$$Tr\{I_z[A_{Q,2},[A_{Q,2}^+,I_z]]\} = \sin^2\theta \, \frac{Tr[(I_+I_-)^2 + 2I_-I_+I_z]}{Tr[I_z^2]} \tag{C2}$$

$$Tr\{I_z[A_{Q,3},[A_{Q,3}^+,I_z]]\} = 4\cos^2\theta \, \frac{Tr[(I_+I_-)^2 + 2I_-I_+I_z]}{Tr[I_z^2]} \tag{C3}$$

The calculation proceeds using the following relations, where $m$ is the quantum number of $I_z$

$$I_\pm I_\mp |m\rangle = [I(I+1) - m(m\mp1)]|m\rangle \tag{C4}$$

$$Tr(I_z^2) = \frac{1}{3} I(I+1)(2I+1) \tag{C5}$$

$$Tr(I_z^4) = \frac{1}{5} I(I+1)(2I+1)\left[I(I+1) - \frac{1}{3}\right] \tag{C6}$$

and gives the results shown in Eq. (17) and Eq. (18).

Appendix D : Calculation of the rate $\Gamma_t$ of donor occupation

Complete calculation of $\Gamma_t$ requires considering the kinetics of generation and recombination for the conduction band, the valence band and the donor and acceptor levels. Although tractable, such calculation leads to intricate results. We assume here, for simplicity and for illustration purposes, that the kinetics of generation and recombination of acceptor levels and of valence holes are similar. Such assumption is reasonable because donor-acceptor recombination, which is specific to holes trapped at shallow acceptors, is known to be less efficient than band to band or exciton recombination.[50] As a result, we consider only one hole specie, for which the total concentration $p$ is the sum of those of valence holes and of neutral acceptors. In steady-state, the rate equations for the concentrations $n_f$ of free electrons and $\Gamma_t N_D$ of electrons trapped at donors are, respectively

$$0 = g - \sigma_c v(1-\Gamma_t)N_D n_f - \frac{n_f}{\tau_r} \tag{D1}$$

$$0 = \sigma_c v(1-\Gamma_t)N_D n_f - \frac{\Gamma_t N_D}{\tau_r} \tag{D2}$$

Eq. (33) is readily obtained using Eq. (D2). The recombination time $\tau_r$ of free and trapped electrons is given by

$$\frac{1}{\tau_r} = k\,p \tag{D3}$$

where $k$ describes the bimolecular electron-hole recombination. Writing further that the total concentrations of photocreated holes and electrons are equal, one obtains successively

$$n_f + \Gamma_t N_D = p - N_A \tag{D4}$$

$$g = n_f \{[\sigma_c v(1-\Gamma_t) + k\Gamma_t]N_D + k[n_f + N_A]\} \tag{D5}$$

$$n_f = (\Gamma_t N_D + N_A)\frac{k\Gamma_t}{\sigma_c v(1-\Gamma_t) - k\Gamma_t} \tag{D6}$$

The excitation power density $P$ corresponding to a given value of $g$ is given by

$$P = gLh\upsilon \tag{D7}$$

where $L$ is the electron diffusion length and $h\upsilon$ is the photon energy. The latter equation assumes that the diffusion length is larger than optical absorption length, and that the surface recombination velocity is negligible. Eq. (46) is readily obtained using Eq. (D5), Eq. (D6), and Eq. (D7) and assuming that $N_D << N_A$.

Acknowledgements

We are grateful to B. Urbaszek, X. Marie and D. Petit for useful discussions and to A. C. H. Rowe for a critical reading of the manuscript.

Figure captions

Fig. 1: Radial dependence of the normalized nuclear polarization, defined by Eq. (21), along the magnetic field direction, (a) and along the perpendicular direction,(b) as a function of distance. Also shown is the radial dependence of the angular average of the nuclear polarization, defined by Eq. (35). The relative magnitude $f_0$ of hyperfine and quadrupolar relaxations, given by Eq. (30), is taken as equal to $10^{-2}$. The distance at which the magnetization is equal to 0.5 is of 0.25 $a_0^*$ in the direction of the magnetic field (a) and 0.45 $a_0^*$ in the perpendicular direction (b) and $\rho_Q = 0.35\, a_0^*$ after angular averaging.(c)

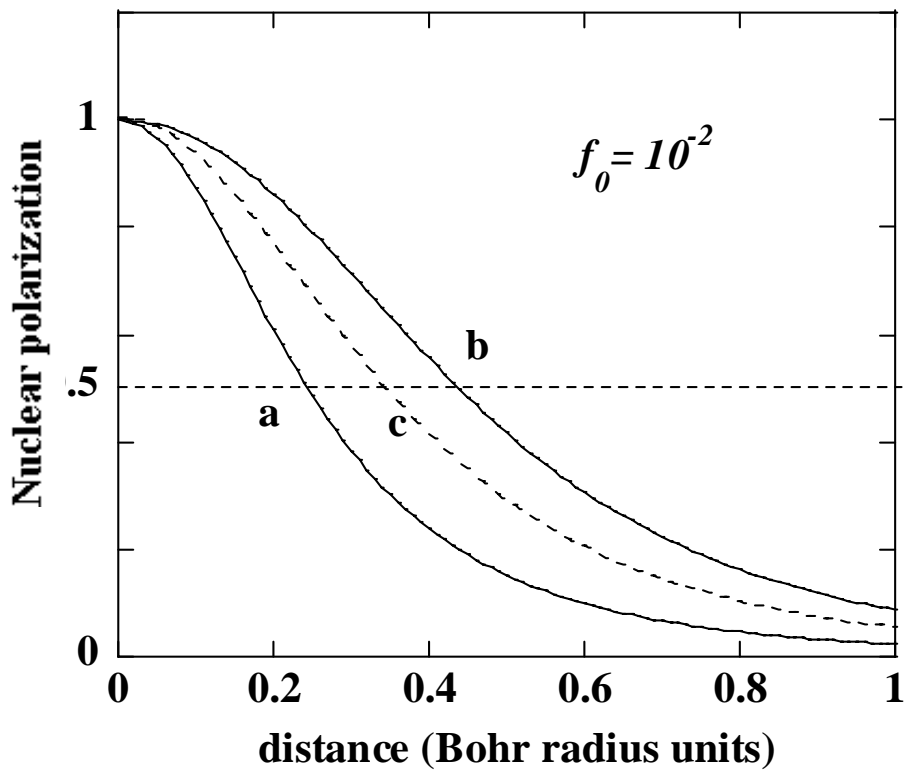

Fig. 2: Dependence of the quadrupolar radius $\rho_Q$ and of $s(\rho_Q)$, which expresses the quadrupolar-induced nuclear field decrease, on the relative magnitude $f_0$ of hyperfine and quadrupolar relaxations. If no light-induced quadrupolar relaxation is present, the quadrupolar radius is replaced by the usual diffusion radius, $\rho_D$, of the order of the Bohr radius. For $f_0 = 10^{-2}$, the quadrupolar radius is $0.35 a_0^*$, and the nuclear field is further decreased by about one order of magnitude.

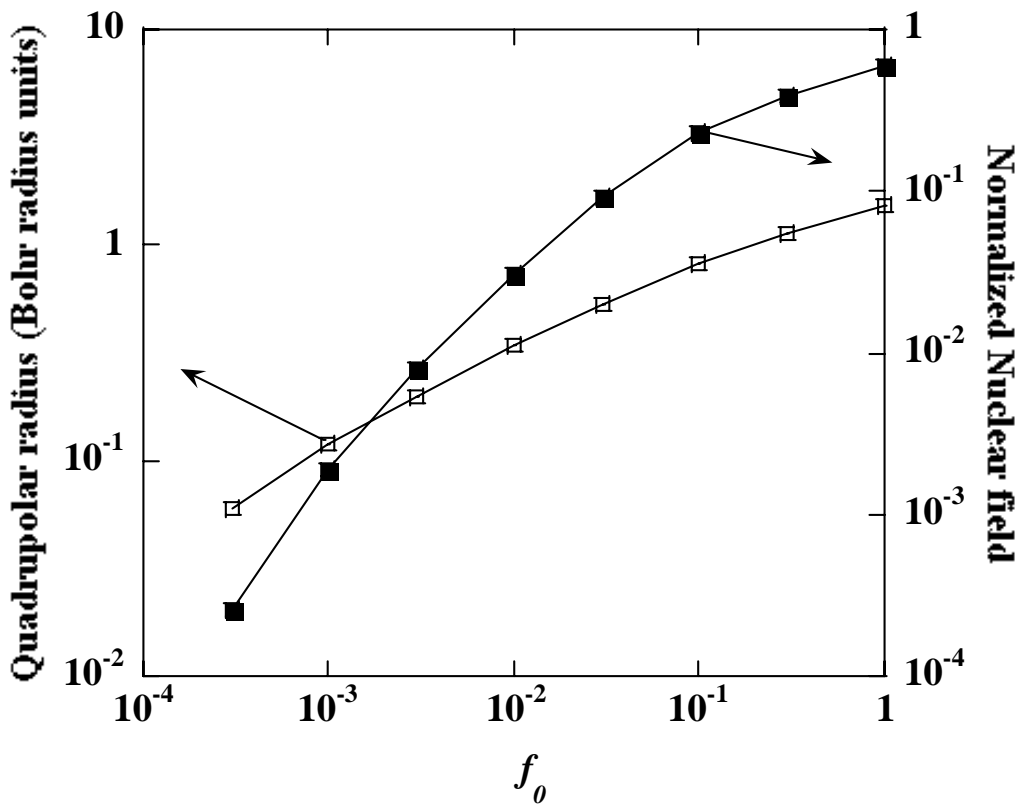

Fig. 3: For $As^{75}$ in GaAs, dependence of the nuclear field on light excitation. The latter quantity is expressed in units of $P_0$ given by Eq. (47), which depends on doping. The donor and acceptor concentrations are taken as $10^{22}$ m$^{-3}$ and $5 \times 10^{22}$ m$^{-3}$, respectively. Curve a shows the rate $\Gamma_t$ of donor occupation. Curve b shows the variation of the quantity $s(\rho_Q)$ which expresses the quadrupolar-induced decrease of the nuclear field with respect to its maximum value, estimated using Ref. (28), set by the presence of spin diffusion. Curve c shows the free electron concentration in units of $N_A$. Curve d shows the dependence of the reduced nuclear field $\alpha_n$, given by Eq. (50), which further takes into account the decrease caused by spin exchange between free and trapped electrons. The hatched area marks the zone where the present model is not valid because of spin diffusion.(see Sec. IIIC)

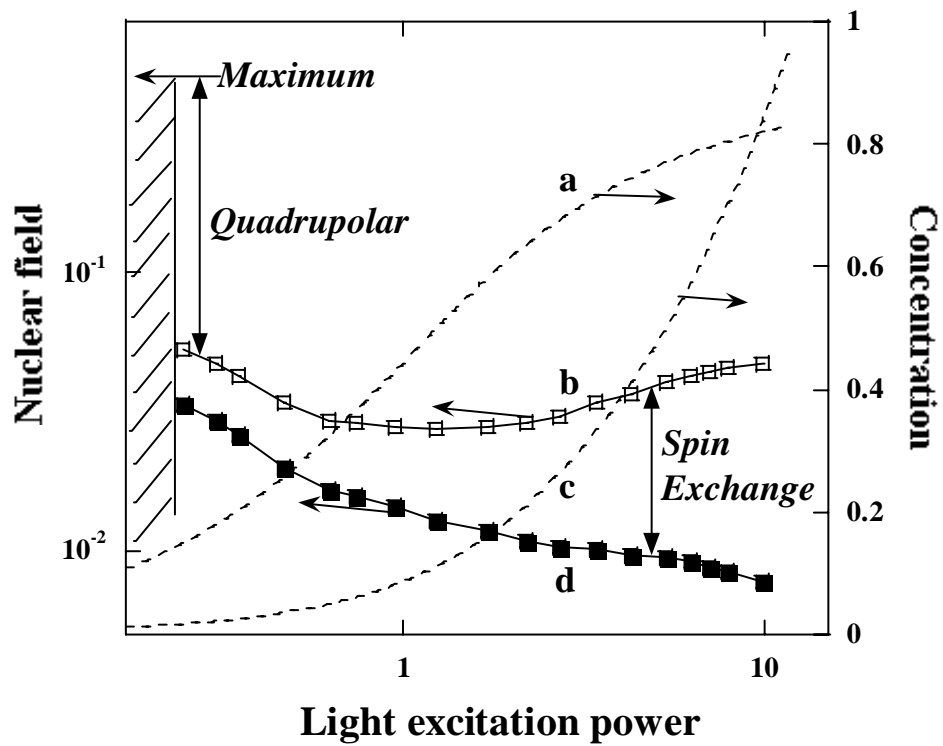

Fig. 4 : Geometry of the magnetic field and electric field configurations. For clarity, the X (X') axis, which lie in the zZ (z'Z') plane and the Y(Y') axis, which lie perpendicular to this plane, have been omitted.

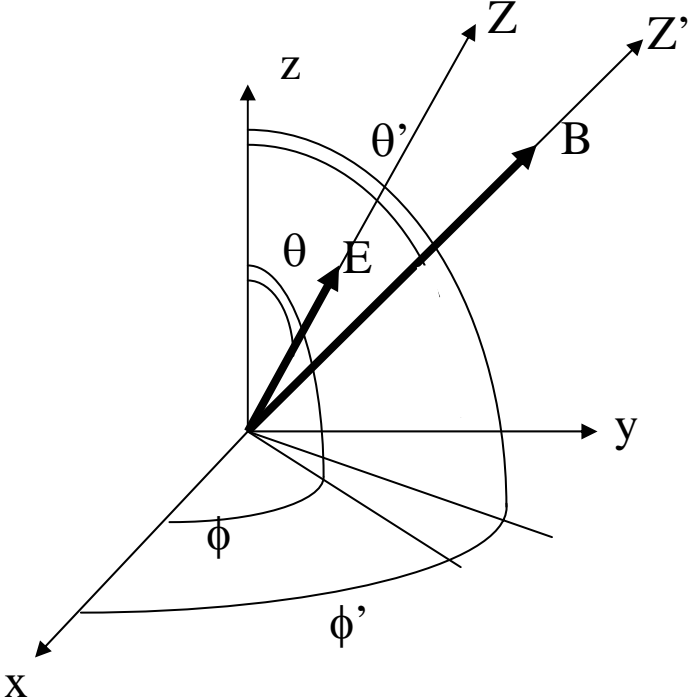

**Table I : Estimate of $b_Q$**

The quantity $b_Q$, which has the dimension of the ratio of a magnetic field to an electric field, is given by Eq. (7) and characterizes the strength of the quadrupolar relaxation. This quantity estimated in Appendix A, is given below for several nucleus/semiconductor matrix combinations (the isotopic specie under consideration is indicated in bold).

| Nucleus | $R_{14}$ ($10^{12}$ m$^{-1}$) | $b_Q$ ($10^{-10}$ Tm/V) |
|---|---|---|
| Ga**As**$^{75}$ | 3.2 | 2.8 |
| **Ga**$^{69}$As | 2.8 | 2.0 |
| **Ga**$^{71}$As | 2.8 | 1.9 |
| **In**$^{115}$As | 4.4 | 0.7 |
| In**As**$^{75}$ | 1.9 | 1.6 |
| **Ga**$^{69}$Sb | 0.7 | 0.51 |
| Ga**Sb**$^{121}$ | 1.9 | 1.2 |
| **In**$^{115}$P | ~ 4 | ~ 0.60 |